\documentclass[aps,prl,showpacs,twocolumn,preprintnumbers,superscriptaddress]{revtex4-1}
\usepackage{amsmath,amssymb}%
\usepackage{graphicx}
\usepackage{dcolumn}
\usepackage{bm}
\usepackage{gensymb}

\begin{document}
\title{Macroscopic evidence of skyrmion lattice inhomogeneity and magnetic vortex states 
in the MnSi $A$-phase}
\date{\today}

\author{S.~{}V.~{}Demishev}
\affiliation {Prokhorov General Physics Institute of RAS, 38 Vavilov str.,  119991 Moscow, Russia}
\affiliation{Moscow Institute of Physics and Technology , 9 Institutsky lane, 141700 Dolgoprudny, Moscow region, Russia}

\author{I.~{}I.~{}Lobanova}
\affiliation{Moscow Institute of Physics and Technology , 9 Institutsky lane, 141700 Dolgoprudny, Moscow region, Russia}

\author{N.~{}E.~{}Sluchanko}
\affiliation {Prokhorov General Physics Institute of RAS, 38 Vavilov str.,  119991 Moscow, Russia} 

\author{V.~{}V.~{}Glushkov}
\email{glushkov@lt.gpi.ru}
\affiliation {Prokhorov General Physics Institute of RAS, 38 Vavilov str.,  119991 Moscow, Russia}
\affiliation{Moscow Institute of Physics and Technology , 9 Institutsky lane, 141700 Dolgoprudny, Moscow region, Russia}
\begin{abstract}
 The magnetic inhomogeneity of the $A$-phase in MnSi chiral magnet is identified for the first time from the precise measurements of transverse magnetoresistance (MR) anisotropy. The area inside the $A$-phase ($A$-phase core) corresponds to isotropic MR having no confinement to the MnSi crystal lattice. \textit{Per contra}, the MR becomes anisotropic both on the border of the $A$-phase and in other magnetic phases, the strongest magnetic scattering being observed when external magnetic field applied along [001] or [00$\bar{1}$] directions. We argue here that the established MR features prove the presence of two different types of the skyrmion lattices inside the $A$-phase, and the dense skyrmion state of the $A$-phase core is built from individual skyrmions similar to Abrikosov-type magnetic vortexes.
 
 \end{abstract}

\pacs{72.15.Gd; 75.25.-j; 75.30.Kz; 73.61.At}

\maketitle

Identifying of universal features in cooperative phenomena induced in quantized vortex matter of different origins ($^3$He superfluid, triplet superconductors and chiral magnets)  stays a real challenge for the physical community \cite{Pfleiderer10,Volovik,Golov}. More than 40 years ago Dudko \textit{et al}. discovered an analogy between the magnetization of type-II superconducting mixed state and antiferromagnet close to the magnetic field induced transition between antiferromagnetic and paramagnetic phases \cite{Dudko72,Dudko75}. Taking FeCO$_3$ as an experimental example, these authors were the first who calculated vortex-like state in the plane perpendicular to the external magnetic field appeared due to the negative surface (domain wall) energy as a result of the specific character of exchange interactions \cite{Dudko75}. Later on, Bogdanov and Yablonskii pointed out that such a \textquotedblleft mixed state\textquotedblright of various magnets may exist due to symmetry factors staying thermodynamically stable in the systems without inversion center \cite{Bogdanov89a,Bogdanov89b}. At that time this mixed state of magnets was definitely considered as a very similar or even perfect analogue of Abrikosov vortexes lattice \cite{Dudko72,Dudko75,Bogdanov89a,Bogdanov89b}.

	Nowadays magnetic vortexes in non-centrosymmetric magnets are generally considered as skyrmions, which are topologically stable knots in the vector field describing magnetization distribution inside the sample \cite{Pfleiderer10}. Historically, manganese monosilicide, MnSi, played an outstanding role in establishing of the skyrmion physics in magnets. Magnetic order, which develops in this material for $T<T_c\sim29$~K, corresponds to a spiral state with the long pitch $\sim18$~nm exceeding essentially the size of the unit cell \cite{Ishikawa76}. This type of magnetic structure is formed under the competition of the strong ferromagnetic exchange $J\sim2.5$~meV and weak Dzyaloshinskii-Moriya (DM) interaction $D\sim0.4$~meV \cite{Grigoriev09}. The skyrmion lattice (SL) state is believed to be hosted by a small pocket on the $B-T$ magnetic phase diagram ($A$-phase), existing in the vicinity of $T_c$ at moderate magnetic fields \cite{Kadowaki82,Muehlbauer12,Neubauer09}. The SL interpretation of the $A$-phase was first deduced both from neutron scattering experiments \cite{Muehlbauer12} and observation of the topological Hall effect \cite{Neubauer09}. These findings were recently confirmed by direct observation of the SL in Lorentz TEM, firstly in thin layers of the MnSi obtained from the bulk crystals \cite{Tonomura12} and later on in the MnSi epitaxial films \cite{Li13}. In theory, the skyrmion phase in MnSi may be treated as compromise between the DM and Zeeman energies. Qualitatively, the skyrmion keeps the spiral configuration of spins inside and ferromagnetic one outside, which results in gaining in both DM and magnetic field energies \cite{Han10}.

	However, the skyrmionic origin of the $A$-phase in MnSi still remains a subject of debates. First of all, the existing theory proves the skyrmion-based states to be stable only in two-dimensional case, and additional mechanisms should be involved to stabilize the SL state in bulk MnSi \cite{Muehlbauer12,Han10,Roessler06,W14}. The surprising disappearance of the $A$-phase recently observed in MnSi epilayers for the out-of-plane external magnetic field leads authors of Ref.~\onlinecite{W14} to a conclusion that the only thermodynamically stables phases on the $B-T$ phase diagram are the conical (C) phase, ferromagnetic (or spin-polarized, SP) phase and paramagnetic (PM) phase and the $A$-phase is completely metastable in 3D case. These results were disputed in\cite{Yu15}, where a comprehensive Lorentz TEM study of the MnSi thin plates of various thicknesses prepared from the single crystals showed the presence of the SL states regardless the crystal orientation with respect to the field. At the same time both Ref.~\onlinecite{W14} and Ref.~\onlinecite{Yu15} agree on the valuable contribution of the magnetic anisotropy and thermal fluctuation effects for the skyrmion stability.

	Note here that the Lorentz TEM technique is applicable to thin quasi-2D samples only, and consequently the SL state should be experimentally proved in the bulk samples. Moreover, the size of the SL pocket on the magnetic phase diagram of MnSi shrinks noticeably when the sample size increases\cite{Yu15}. In this situation, new experiments based on the different methods are desirable to elucidate the origin of the $A$-phase in the 3D magnet. At the same time, even if skyrmionic interpretation is assumed, the situation remains controversial. Grigoriev \textit{at al}. pointed out that in current literature the term \textquotedblleft skyrmion lattice\textquotedblright for the $A$-phase in MnSi is used for marking of two mutually exclusive physical situations \cite{Grigoriev14}. The first one is very close to the initial concept of Abrikosov-type magnetic vortexes. According to the basic study \cite{Roessler06} the skyrmions are real quasi-particles, which may build hexagonal SL with the period not necessary related to the spiral pitch, although it should be of the same order of magnitude \cite{Han10}. Apparently, in this concept the melting of the SL into smaller skyrmion-based clusters may be expected as well as skyrmion-like magnetic fluctuations above $T_c$ \cite{Pappas09}. The second concept originates from the $A$-phase model based on the triple-\textbf{k} spin structure stabilized by Gaussian thermal fluctuations under magnetic field \cite{Muehlbauer12}. The latter ansatz accounting skyrmion-like topology, which includes protected knots and windings in the $A$-phase magnetic structure \cite{Muehlbauer12,Neubauer09}, and therefore this type of SL is nothing but a complicated magnetic phase, which could not decay into individual skyrmions. The search for the SL melting effects performed in \cite{Grigoriev14} provided negative result, leading to the conclusion that quasi-particle skyrmion scenario of the $A$-phase is not correct.

	In the present work, we report the result of the simple experiment consisting in an investigation of the angular dependences of the magnetoresistance (MR) inside and near the $A$-phase $B-T$ domain in the bulk MnSi single crystals. The magnetic scattering of charge carriers is known to dominate in MnSi under negligible Boltzmann contribution to MR \cite{Demishev12}, so that this physical parameter happens to be very sensitive to the magnetic structure. Therefore studying of the MR anisotropy may bring additional information, which is hardly accessible in direct structural studies. The analysis of the MR data allows concluding that $A$-phase is not magnetically homogeneous and most probably includes the region of the magnetic vortex state constructed of individual skyrmions. The obtained structure of the $A$-phase was missed in all previous studies and should provide a new look on the problems of skyrmion stability in the 3D case, melting of the skyrmion lattice and alternative skyrmion-based interpretations of the $A$-phase nature.
	
\begin{figure}[t]
		\includegraphics[width=0.7\linewidth]{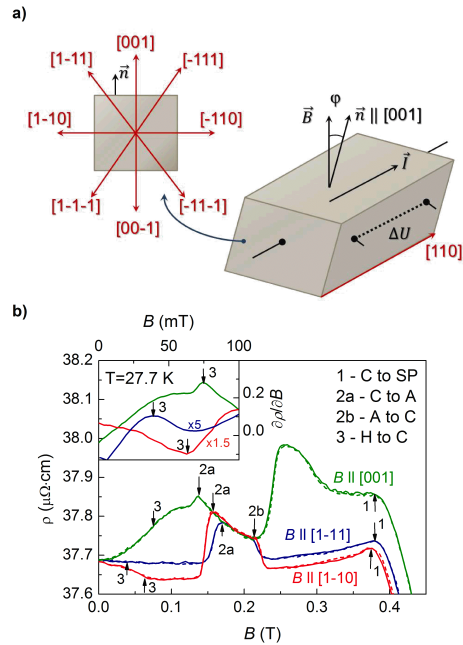}
		\caption{\label{fig1}(Colour on-line) (a) Experimental geometry for the transverse MR anisotropy measurements. (b) Field dependences of the resistivity as measured at $T=27.7$~K for magnetic fields applied along different crystallographic directions. Arrows denote magnetic transition between the cone (C) and spin-polarized (SP) phases (1), the segment of the $A$-phase stability (2~a,b) and the phase boundary between the helix (H) and C phases (3). Inset in panel~b shows the $\partial\rho/\partial{B}$ data in the units of 10$^{-5}$ $\Omega\cdot${cm}/T}
	\end{figure}

	High quality single crystals of MnSi identical to those studied earlier in \cite{Demishev12} were investigated here. Details concerning samples preparation and characterization can be found elsewhere \cite{Demishev12}. For MR measurements, the DC excitation current was applied 
	parallel to [110] direction in all cases. The step motor driven installation allowed step-by-step rotating of the sample ($\Delta\varphi$$\sim$1.8$\degree$) in a steady magnetic field supplied by a superconducting solenoid. The rotation of the single crystal around the [110] axis allowed changing direction of magnetic field \textbf{B} in the transverse MR experiment so that vector \textbf{B} can gradually pass in one run the main crystallographic directions [001], [1$\bar{1}$1], [1$\bar{1}$0], [111], and respectively [00$\bar{1}$], [$\bar{1}$1$\bar{1}$], [$\bar{1}$10], [$\bar{1}$11]. The position of the crystal is identified by the angle $\varphi$ between the vectors \textbf{B} and [001] (Fig.~{}\ref{fig1},a). The standard four-probe technique was used to measure sample resistivity $\rho(B,T)$ as a function of temperature and magnetic field, the relative accuracy of the resistivity data was about $\sim$0-5 and the accuracy of temperature stabilization in the diapason 26-30~K reached $\sim$2~mK independent on the angle $\varphi$ and magnetic field magnitude up to 0.5~T.
	 
\begin{figure}[t]
	\includegraphics[width=0.7\linewidth]{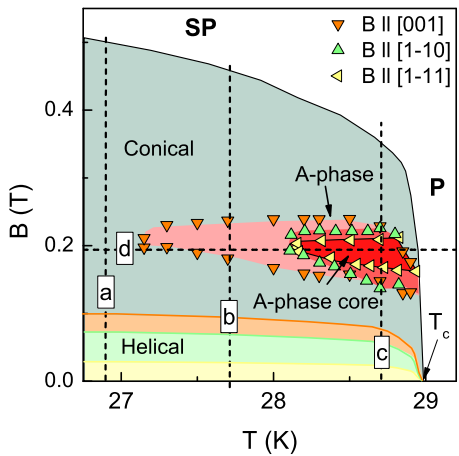}
	\caption{\label{fig2}(Colour on-line) The magnetic phase diagram of MnSi reestablished from the MR measurements.  Dashed lines show different cuts taken at (a) $T=26.9$~K, (b) $T=27.7$~K, (c) $T=28.7$~K and (d) $B=0.194$~T.}
	\end{figure}

	The field dependences of the MR for magnetic fields applied along the main crystallographic directions can be used for determination of the magnetic phase diagram in the MnSi studied sample \cite{Kadowaki82}. The typical $\rho(B,T=$ const) data are presented in Fig.~\ref{fig1},b for $T=27.7$~K. The most pronounced kinks correspond to magnetic transitions between cone (C) and spin-polarized (SP) phases and to the phase boundaries of the $A$-phase (denoted as 1 and 2~a,b, respectively). The transition 3 between helix (H) and C phases does not influence significantly on the $\rho(B,T=$ const) curves but becomes visible as a maximum or minimum of the derivative $\partial\rho/\partial{B}$ (inset in Fig.~\ref{fig1},b). The analysis of the $\rho(B,T)$ field dependences allowed obtaining the magnetic phase diagram (Fig.~\ref{fig2}). It is obvious that the largest pocket of the $A$-phase corresponds to the case \textbf{B}$\parallel$[001]. Comparison with the known magnetic phase diagrams for MnSi extracted from the magnetotransport and magnetization measurements \cite{Kadowaki82,Muehlbauer12,Neubauer09,Bauer12} demonstrates a reasonably good agreement of the present results (Fig.~{}\ref{fig2}) with those ones, which were reported previously. 
	
	According to suppositions made in Ref. {}\onlinecite{W14}, not only the kinks on the $\rho(B)$ curves but also magnetic hysteresis is an essential fingerprint of the $A$-phase $B-T$ region on the magnetic phase diagram. This statement is based on the pioneering results \cite{Kadowaki82}, where strong MR hysteresis for the up and down sweeps of the magnetic field was found. 	In view of the proposed elimination of the $A$-phase from the set of \textquotedblleft true\textquotedblright magnetic phases due to its metastable nature \cite{W14} 	the hysteresis effects may be considered as an attribute of metastability. 
	In Fig.{}\ref{fig1}b 	the accuracy of the temperature stabilization during field scans was 
	$\sim$3~mK and the $\rho(B)$ 	data for sweeping up (solid lines) and sweeping down (dashed lines) almost coincide. When the stabilization level was increased to $\sim$5~mK 
	the magnetic hysteresis became more pronounced.	These observations clearly show that hysteresis 	features may be completely attributed to the thermal effects occuring 
	at the $A$-phase boundaries. Indeed, from the theoretical point of view the transitions between the C phase and $A$-phase should be the first order ones \cite{Han10}. Therefore, the $A$-phase in MnSi has well-defined phase boundaries similar to another phases on the $B-T$ diagram 	and hence possible hysteresis effects \cite{Kadowaki82,W14} 	 are nothing but experimental artefacts caused by the discrepancy between the measured and real sample temperature.

	A remarkable feature of the $\rho(B)$ dependences in Fig.~{}\ref{fig1},b is a coincidence of the resistivity values for three main crystallographic directions in the interval around $0.2$~T. This segment corresponds to the area on the $B-T$ magnetic phase diagram, which is common for the three $A$-phase pockets in Fig.~{}\ref{fig2} (hereafter this area will be denoted as $A$-phase core). At the same time, the $\rho(B)$ curves exhibit strong anisotropy outside of the specific interval (Fig.~{}\ref{fig1},b).

\begin{figure}[t]
	\includegraphics[width=1\linewidth]{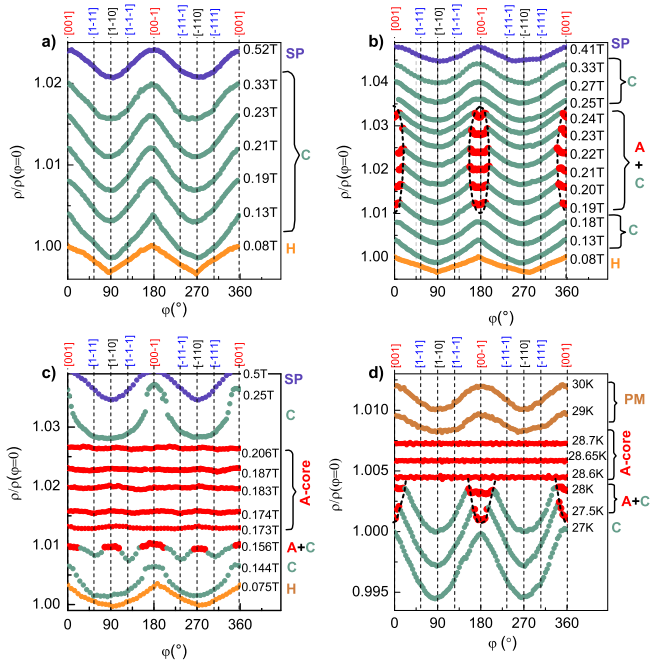}
	\caption{\label{fig3}(Colour on-line) Angular dependences of the MR for the cuts of the $B-T$ magnetic phase diagram at $T=26.9$~K~(no A-phase, a), $T=27.7$~K~(short segments of A-phase, b), $T=28.7$~K~(A-core crossing, c) and $B=0.194$~T~(A-core crossing, d). Bold dashed lines in the panels b and d denote the segment of the A-phase, which appears near \textbf{B}$\parallel$[001] direction. The magnetic phases corresponding to different $\rho(\varphi)$ curves are given at the right of each panel by the same color}
	\end{figure}

	In order to analyze the magnetic anisotropy effects inside and around the $A$-phase, the angular dependences of the resistivity $\rho(\varphi)$ were investigated along different lines on the $B-T$ magnetic phase diagram (dashed lines a-d in Fig.~{}\ref{fig2}). Three cuts were taken at fixed temperatures $T=26.9$~K, $T=27.7$~K, $T=28.9$~K under varying magnetic field (lines a-c in Fig.~{}\ref{fig2}) and one cut was measured at fixed magnetic field $B=0.194$~T under varying temperature (line d in Fig.~{}\ref{fig2}). Fig.~{}\ref{fig3}
	 presents the resultant normalized $\rho(\varphi)/\rho(0)$ data shifted for clarity along $y$ axis. The isotherm $T=26.9$~K lying outside of the $A$-phase (section a in Fig.~{}\ref{fig2}) captures the sequence of the magnetic phase transitions H$\rightarrow$C$\rightarrow$SP. It is visible (Fig.~{}\ref{fig3},a) that the direction of the strongest magnetic scattering \textbf{B}$\parallel$[001] leading to the maximal resistivity (or equally \textbf{B}$\parallel$$[00\bar{1}]$) is kept in all magnetic phases although the shape of the $\rho(\varphi)$ curves varies. 
	
Entering into the $A$-phase region for \textbf{B}$\parallel$[001] except the $A$-phase core ($T=27.7$~K, line~b in Fig.~{}\ref{fig2}) results in deviation from the above behavior. In the interval of the magnetic fields corresponding to the $A$-phase, additional kinks on the MR angular dependences appear and the maximal magnetic scattering is observed within $\Delta\varphi(B)\sim\pm20\degree$ from [001] or [00$\bar{1}$] axes (dashed lines in Fig.~{}\ref{fig3},b). Inside the $\Delta\varphi(B)$ interval $\rho(\varphi)\approx$ const within experimental accuracy (Fig.~{}\ref{fig3},b). Outside the diapason corresponding to the $A$-phase boundaries for \textbf{B}$\parallel$[001], the character of the $\rho(\varphi)$ curves evolution with magnetic field is conserved. Therefore, the comparison of the MR data in Figs.~{}\ref{fig3},a and~{}\ref{fig3},b suggests that the \textquotedblleft$A$-phase for \textbf{B}$\parallel$ [001]\textquotedblright may exist in a range of angles $\Delta\varphi(B)$ around [001] marked by dashed lines in Fig.~{}\ref{fig3},b), so that $\Delta\varphi(B)\rightarrow0$ at the phase boundaries. Consequently, the total angular dependences of the MR $\rho(\varphi)$, where transitions into $A$-phase are observed, result from contributions of both the C phase and the $A$-phase pinned around the \textbf{B}$\parallel$[001] (or \textbf{B}$\parallel$[00$\bar{1}$]) direction.

Cuts $T=28.7$~K and $B=0.194$~T passing through the $A$-phase core (lines c and d in Fig.~{}\ref{fig2}) demonstrate dramatically different behavior. Inside the $A$-phase core, the resistivity becomes isotropic and $\rho(\varphi)=$ const for any angle (Figs.~{}\ref{fig3},c and~{}\ref{fig3},d), this characteristic feature not depending on the section type ($T=$ const or $B=$ const). Hence the angular range of the $A$-phase core stability is not confined by limited $\Delta\varphi(B)$ interval along any crystallographic direction. Outside this exclusive region on the magnetic phase diagram, the variation of the resistivity angular  dependences is qualitatively the same as described above (Fig.~{}\ref{fig3}). It is worth noting that the direction of the maximal magnetic scattering \textbf{B}$\parallel$[001] remains the same in both SP and PM phases  (Figs.~{}\ref{fig3},a and~\ref{fig3},d).

The qualitative change of the magnetic scattering anisotropy in the $A$-phase core with respect to all other areas on the $B-T$ magnetic phase diagram provides clear evidence that the nature of this phase in MnSi is completely different. Regardless of whether this phase is metastable or not, its rotational symmetry has no confinement to any crystallographic directions. Moreover, the data in Fig.~{}\ref{fig3} show unambiguously that $A$-phase is not homogeneous. Indeed, when considering route $B=$const (Fig.~{}\ref{fig3},d) it is visible that increase of temperature results at first in entering into the $A$-phase along \textbf{B}$\parallel$[001], which exists in small range $\Delta\varphi(T)$ around [001] or [00$\bar{1}$] (dashed lines in Fig.~{}\ref{fig3},d), and the respective $\rho$($\varphi$) curves reflect the superposition of the C phase and $A$-phase segments (similar to that one in Fig.~{}\ref{fig3},b). When the temperature reaches the interval corresponding to the $A$-phase core, the resistivity does not depend on the angle and consequently at the boundary between the $A$-phase and $A$-phase core the symmetry of the MR anisotropy undergoes qualitative change at the boundary between the $A$-phase and $A$-phase core. Intermediate $\rho(\varphi)$ behavior inside the $A$-phase, which precedes the onset of the $\rho(\varphi)=$ const in the $A$-phase core, is also evident from Fig.~{}\ref{fig3},c (see the data for $B=0.156$~T). However in the latter case the angular variation of the resistivity in the $A$-phase becomes more complicated due to the influence of close phase boundaries between the C phase and $A$-phases for \textbf{B}$\parallel$[001], \textbf{B}$\parallel$[1$\bar{1}$0] and \textbf{B}$\parallel$[1$\bar{1}$1] (Fig.~{}\ref{fig2}). It is worth noting that the first indications of the $A$-phase magnetic inhomogeneity were reported in the pioneering work \cite{Kadowaki82}. However, only the analysis of the MR angular dependences allows proving this statement.

	Results of the present work show that MnSi \textquotedblleft forgets\textquotedblright any magnetic anisotropy related to the crystalline lattice inside the A-phase core. This fact can be easily recognized within the theory of the mixed state of magnets to be analogous to a mixed state of superconductors \cite{Dudko72,Dudko75,Bogdanov89a,Bogdanov89b}. Indeed, the Abrikosov-like magnetic vortexes are superstructural entities, which are linked to the direction of magnetic field rather than to any axes in crystal. Therefore the observation of the $\rho(\varphi)=$ const looks reasonable for a state built on the magnetic vortexes similar to those in type II superconductors. Nevertheless, numerous experimental and theoretical evidences available to date in the skyrmion physics require more detailed analysis of the experimental facts obtained in the present study.

	First of all, it is necessary to mark that the case where SL is not coupled to crystallographic directions is this one of SL constructed from individual skyrmions, for which a direct analogy with the vortex state of a superconductor can be proven mathematically \cite{Han10}. At the same time, the triple-\textbf{k} spin configuration based models should imply some anisotropy effects as long as this ansatz to SL description must include an explanation of the anisotropic boundaries of the $A$-phase appearing in the highly symmetric cubic lattice (Fig.~{}\ref{fig2}). Secondly, the analysis of the MR angular dependences strongly supports the inhomogeneous structure of the $A$-phase, where the isotropic $A$-phase core is surrounded by some anisotropic magnetic phase. Than, following the Occam's razor principle, it is possible to suggest a hypothesis for the description of the $A$-phase structure, where isotropic SL state built from individual skyrmions as quasiparticles corresponds to $A$-phase core, which, in turn, is surrounded by the anisotropic triple-\textbf{k} based skyrmion lattice. Apparently in this model no melting to individual skyrmions is possible as long as skyrmions are confined inside another type of the SL, which is unable to decay into individual quasiparticles. Therefore the problem pointed out by Grigoriev \textit{et al}. \cite{Grigoriev14} gets a natural solution: instead of the rivalry the two different SL phases coexist inside the $A$-phase.

	The above consideration shows that the study of the skyrmion-based states stability in MnSi performed to date \cite{Pfleiderer10,Muehlbauer12,Han10,Roessler06,Yu15,W14} is not complete, because the thermodynamic stability of the SL was analyzed with respect to the C and SP (ferromagnetic) phases. The complete analysis, which has not been done so far, should include evaluation of the energies of the dense skyrmion phases of various types, which differ by their magnetic anisotropy. Prior to such calculations it is not possible to come to a definite conclusion concerning the stability in 3D case of the SL states, which consist of the individual skyrmions. Moreover, there are no reasons to consider this phase as a metastable one as long as a very reasonable stabilization mechanism in the 3D case is included in the triple-\textbf{k} model \cite{Muehlbauer12}, which is likely responsible for the phase boundaries between the C phase and $A$-phase. It may turn out that some aspects of metastability may be relevant to the 2D case, where both expansion and disappearance of the $A$-phase domain can be observed \cite{Tonomura12,Li13,W14,Yu15}, rather than to the 3D case.

	Summarizing up, the study of the MR angular dependences in MnSi reveals the magnetic inhomogeneity of the $A$-phase. The MR becomes isotropic  in the area inside the $A$-phase, which is common to all crystallographic directions ($A$-phase core), whereas both in the surrounding part of the $A$-phase and in other magnetic phases the magnetoresistance is anisotropic with the strongest magnetic scattering occurring under external magnetic field aligned along [001], or [00$\bar{1}$] directions. Analysis of the experimental data strongly suggests a magnetic transition between different types of the skyrmion lattices inside the $A$-phase, where dense skyrmion state of the $A$-phase core is built from individual skyrmions analogously to Abrikosov-type magnetic vortexes.

The authors are grateful to S.M.~Stishov for stimulating discussions at the early stage of this work. This work was supported by Programmes of Russian Academy of Sciences \textquotedblleft Electron spin resonance, spin-dependent electronic effects and spin technologies\textquotedblright, \textquotedblleft Electron correlations in strongly interacting systems\textquotedblright and by RFBR grant 13-02-00160.

\end{document}